# Emergence of novel hydrogen chlorides under high pressure

Qingfeng Zeng[a, f, *], Shuyin Yu[a, f], Duan Li[a, f], Gilles Frapper[b, *] and Artem R. Oganov[c, a, d, e, *]

[a]*International Center for Materials Discovery, School of Materials Science and Engineering, Northwestern Polytechnical University, Xi'an, Shaanxi 710072, PR China*

[b]*IC2MP UMR 7285, University of Poitiers - CNRS, Poitiers 86022, France*

[c]*Skolkovo Institute of Science and Technology, Skolkovo Innovation Center, Bldg. 5, Moscow 143026, Russia*

[d]*Moscow Institute of Physics and Technology, Dolgoprudny, Moscow Region 141700, Russia*

[e]*Department of Geosciences, Center for Materials by Design, and Institute for Advanced Computational Science, State University of New York, Stony Brook, NY 11794-2100, USA*

[f]*Science and Technology on Thermostructural Composite Materials Laboratory, School of Materials Science and Engineering, Northwestern Polytechnical University, Xi'an, Shaanxi 710072, PR China*

∗The author to whom the correspondence should be addressed to: Qingfeng Zeng, email: qfzeng@nwpu.edu.cn; Artem R. Oganov, e-mail: artem.oganov@stonybrook.edu; Gilles Frapper, e-mail: gilles.frapper@univ-poitiers.fr

**Abstract**

HCl, a 'textbook' example of a polar covalent molecule, is a well-known compound of hydrogen and chlorine. Inspired by the discovery of unexpected stable stoichiometries of sodium chlorides, we performed systematic searches for all stable compounds in the H-Cl system from ambient pressure to higher pressures up to 500 GPa using variable-composition *ab initio* evolutionary algorithm USPEX. We found several compounds that are stable under pressure, i.e. HCl, $H_2Cl$, $H_3Cl$, $H_5Cl$ and $H_4Cl_7$, which display a rich variety of chemical bonding types. At ambient pressure, $H_2$, $Cl_2$ and HCl molecular crystals are formed by weak intermolecular van der Waals interactions and adjacent HCl molecules connect with each other to form asymmetric zigzag chains, which become symmetric under high pressure. In hydrogen-rich chlorides, $H_2$ and HCl react to form the thermodynamically stable $H_3Cl$ crystalline compound in which molecular cyclic $H_3^+$ cations are stabilised by the $Cl^-$ sublattice. Increasing the amount of hydrogen leads to stable solid-state $H_5Cl$, in which $H_2$ formally combines with $H_3^+$ to form $H_5^+$ cations. Additionally, chlorine-based Kagomé layers are formed with intercalated zigzag HCl chains in chlorine-rich hydrides. These discoveries help to understand how varied bonding features can co-exist and evolve in one compound under extreme conditions.

## 1. Introduction

In 1935, Wigner[1] proposed an insulator-to-metal transition in solid hydrogen under pressure. The dissociation, rebonding and polymeriztion, and eventual metalization of molecular systems under high pressure provide conditions for producing novel physical and chemical phenomena. Chemical bonding is the key to understanding the structure and properties of materials, and impacts a broad range of fields including physics, chemistry, biology, materials and earth sciences. It is known that very unusual and unexpected compounds can become stable under pressure, and the chemical nature of many of these compounds are still not understood[2,3]. For example, under pressure, the simple Na-Cl system contains stable compounds such as $Na_3Cl$, $Na_2Cl$, $Na_3Cl_2$, $NaCl_3$ and $NaCl_7$, as well as 'normal' NaCl[4]. While NaCl is a 'textbook' example of an ionic crystal, HCl is an archetypal polar covalent molecule. One can expect large differences between the Na-Cl and H-Cl systems; however, we can imagine that the HCl system is even more diverse due to its increased variety of intermolecular and intramolecular interactions. In addition, hydrogen adds extra excitement: the high-pressure metallic phase of hydrogen has been predicted to possess a nearly room-temperature superconductivity ($T_c$ ~240K)[5], and while superconducting metallic hydrogen remains elusive, recent calculations using USPEX predicted the high-pressure stability of the compound $H_3S$[6] with a $T_c$ reaching 190-200 K; this prediction was subsequently verified experimentally[7].

It has been suggested[8] that alloying hydrogen with other elements can greatly reduce the pressure required for stability of the superconducting state, and one may think not of only metal hydrides, but also of compounds with more electronegative atoms, such as the aforementioned $H_3S$, or the H-Cl compounds studied here. Several H-Cl compounds were predicted in a recent work, but we still found some results[9-11] are in contradiction with our findings. Besides the structure prediction, the bonding features and superconductivity were not fully explored. Here, we attempt to give an exhaustive picture of the H-Cl system under pressure.

## 2. Methods

Searches for stable compounds in the H-Cl system from ambient pressure up to 500 GPa were performed using the variable-composition *ab initio* evolutionary algorithm USPEX[12,13]. In these variable-composition searches, all H-Cl compositions were allowed, under the constraint that the total number of atoms in the unit cell should not be greater than 30 atoms. The first generation contained 80 candidate structures, while all subsequent generations contained 60 structures. In each new generation, 40% of the structures were produced by heredity, 20% by softmutation, 20% by transmutation and 20% were produced randomly. Local geometry relaxations were performed using density-functional calculations with the help of the VASP code[14]. These calculations are based on the Perdew-Burke-Ernzerhof (PBE) functional[15], which ascribes to the generalised gradient approximation (GGA) level of theory. We used the all-electron projector-augmented wave method (PAW)[16], with an outermost core radii of 1.1a.u. for the H atom and 1.9 a.u. for the Cl atom ([Ne] core). The plane-wave kinetic-energy cutoff was set to 900 eV, and uniform -centred $k$-meshes with a reciprocal space resolution of $2\pi \times 0.06$ Å$^{-1}$ were used for sampling the Brillouin zone. These settings enabled excellent convergence of the energy differences, stress tensors, and structural parameters. Denser $k$-point meshes in reciprocal space with a resolution of $2\pi \times 0.03$ Å$^{-1}$ were used for further calculations.

All compounds presented here were confirmed to be dynamically stable at their predicted pressure ranges of stability (SM, Fig. S2). Phonon calculations were performed using density-functional perturbation theory (DFPT)[17]. Detailed structural information is given in Supplementary Materials (SM, Table S1).

All molecular species were optimised at different levels of theory (B3PW91, PBEPBE, MP2, CCSD(T)). A triple-zeta basis set was employed for all atoms, which was increased using both polarized and diffuse functions, aug-cc-pVTZ[18,19]. For each structure, the analytic Hessian was calculated to obtain the vibrational frequencies and determine the nature of the stationary point (local minimum). Calculations were performed using the Gaussian 09 computational programs[20]. As the results followed the same trend, we present here only the electrostatic potential results based on the B3PW91 functional. The surface potentials, $V_S(r)$, were obtained using the Wave Function Analysis-Surface Analysis Suite (WFA)[21].

## 3. Results and Discussion

Figure 1 shows the convex hulls of the H-Cl system in the pressure range 0-500 GPa. Elemental hydrogen and chlorine in their most stable forms, i.e., the *P6$_3$/m*, *C2/c*, *Cmca*-12, *Cmca* and *I4$_1$/amd* structures for $H_2$[22] and *Cmca*, *Fmm*2-28, *Immm* and *FCC* structures for chlorine[23], were adopted as the reference states in their pressure ranges of their stability. The convex hull is a set of thermodynamically stable states, which are stable with respect to disproportion into other phases or pure elements. Any structure, for which the enthalpy of formation lies on the convex hull, is considered to be thermodynamically stable and - in principle - can be synthesised from any isochemical mixture. Besides reproducing the well-known HCl phase (space group *Cmc*$2_1$), we also found the stable stoichiometries 2:1, 3:1, 5:1 and 4:7. After analysing their crystal structures, we identified a rich variety of structures with symmetric and asymmetric zigzag HCl chains, host-guest inclusion compounds, chlorine Kagomé layers, $H_3^+$ and $H_5^+$ units, interpenetrating graphene-like chlorine nets, and other structures, which will be illustrated in the following text.

The pressure-composition phase diagram of the H-Cl system is shown in Fig. 2. We find that HCl remains a stable compound in the pressure ranges 0-160 GPa and >251 GPa, but decomposes into $H_2Cl$ and $H_4Cl_7$ at the pressures between these ranges. Similar reentrant behaviour is also observed for $H_3Cl$. The stable hydrogen-rich compound $H_3Cl$ emerges on the phase diagram at 19 GPa and decomposes into $H_2Cl$ and $H_5Cl$ at 128 GPa and reappears at 320 GPa. For $H_2Cl$, the monoclinic *C*2/*c* structure appears at 44 GPa and transforms to the highly-symmetric *R*-3*m* structure at 341 GPa. The hydrogen-richest compound is $H_5Cl$. This phase appears at 106 GPa and becomes unstable at 479 GPa. Additionally, we found a chlorine-rich compound, $H_4Cl_7$, which becomes stable at 90 GPa and remains stable up to 445 GPa when it decomposes into a mixture of HCl and Cl.

*3.1. HCl: A brief revisit of the pressure-induced symmetrisation of the zigzag chains*

At ambient conditions, HCl adopts the orthorhombic *Cmc*$2_1$ structure (Fig. 3a), which remains stable up to 39 GPa. In this structure, HCl forms zigzag chains with a H-Cl distance of 1.313 Å, compared with the calculated H-Cl bond length of 1.311 Å in the gas phase (exp. 1.274 Å). The intermolecular H-Cl separations in *Cmc*$2_1$ are 2.219 Å, much less than the sum of the van der Waals radii of 2.950 Å[24] and typical of hydrogen bond.

The structure of the orthorhombic *Cmcm* phase is shown in Fig. 3b. In this condensed phase, HCl-HCl contacts have an L-shaped geometry, with the H-Cl bond axis of one molecule perpendicular to the other one. A similar herringbone pattern is also observed in solid $X_2$ (X, halogen)[25,26]. Fig. 4a illustrates why the HCl chain possesses a zigzag shape (L-shaped geometry) rather than a linear or all-trans one. Electrostatic arguments can be used to find the answer. The computed electrostatic potential $V_S(r)$ on the 0.001 a.u. surface of HCl is displayed in Fig. 4b, computed at the B3PW91/aug-cc-pVTZ level using the WFA surface analysis suite[18-21]. A region of positive electrostatic potential – the so-called σ-hole[27,28] can be seen on the outer surface of the hydrogen, with a maximum positive $V_{s,max}$ value of 44.6 kcal/mol. Due to the two chlorine lone pairs of electrons in the $3p_x$ and $3p_y$ orbitals, the most negative values, $V_{s,min}$, are on the lateral rings of Cl negative $V_s(r)$, shown in red with a $V_{s,max}$ of -9.5 kcal/mol. Thus the hydrogen atom may be considered as a Lewis acid site, and the halogen atom interacts as an electron donor through its π lone pairs, thus the angle for the H-Cl---H arrangement should be around 90°. This was also observed for the *Cmc*$2_1$-HCl structure (Cl---H-Cl angles 174.8-175.7°). Even if the energy of this highly directional non-covalent interaction driven by electrostatic force is small - less than 1.5 kcal per HCl unit in a single (HCl)$_n$ chain – hydrogen bonding governs the crystal architecture of HCl up to 50 GPa. To summarise, this zigzag topology is closely related to the anisotropy of the electron charge distribution of the halogen (Cl) atom as schematically depicted in Fig. 4b. Molecular orbital arguments can also be used to understand this L-shaped coordination, as depicted in Fig. 4c. A stabilising interaction between two H-Cl units occurs in such a manner that a filled $p_\pi(Cl)$ orbital – a lone pair of Cl – is oriented towards the H-Cl level of the other (Fig. 4d).

When pressure increases, the difference between the alternating short and long H-Cl separations becomes smaller. Eventually, this difference disappears and symmetric hydrogen bonds appear - symmetric HCl-based zigzag chains are encountered in several phases at different compositions. Hydrogen bond symmetrisation in HCl under high pressure is well studied, both experimentally and theoretically[11, 29]. In HCl, symmetrisation occurs experimentally at 51 GPa at T = 300 K, which can be attributed to softening of the symmetric stretch $A_1$ mode. A $Cmc2_1 \rightarrow Cmcm$ transformation is predicted to occur at ~40 GPa, then a $Cmcm \rightarrow P2_1/m$ transition occurs above 233 GPa[11]. The $Cmc2_1 \rightarrow Cmcm$ structural transformation is calculated to occur at about 35 GPa, in good agreement with Duan et al.[28]. Recall that HCl experimentally adopts the disordered phase I at room temperature while it requires 19 GPa to transform into phase III ($Cmc2_1$)[30].

In $Cmcm$-HCl at 100 GPa, the symmetric zigzag chains are stacked along the a-axis with an interchain spacing of 1.940 Å and a shortest Cl-Cl separation of 2.883 Å. Delocalized three-center two-electron bonding is expected along the σ H-Cl chains (Fig. 4e), which explains the increase in the covalent H-Cl bond length from 1.313 Å at 0 GPa to 1.442 Å here, and the decrease in the hydrogen-bond separation to 2.219 Å at 0 GPa. In agreement with the VSEPR model, here the zigzag chain contains two-coordinate bent chlorine atoms (H-Cl-H angle of 84.2°) and two-coordinate almost linear hydrogen bonds.

The hydrogen bond symmetrisation process for HCl under high pressure is well characterised[23] and can be considered as a 'textbook' example of a second-order Peierls distortion in weakly one-dimensional covalent dimer-based chains. The symmetric zigzag ($H_2Cl_2$) chain presents Peierls-type instability of the -bands at the Fermi level, which induces structural distortion along the H-Cl backbone – creation of short and long H-Cl bonds – when the pressure is reduced to ambient conditions. This dimerization process leads to the thermodynamically stable covalent HCl monomers and increases the band gap. Finally, in addition to hydrogen bond symmetrisation, reduction of the van der Waals volume occurs under compression, leading to stronger interchains Cl---Cl interactions. From 0 to 100 GPa, the interplane stacking separation decreases drastically from 3.506 Å (0 GPa, $Cmc2_1$, Cl---Cl 4.388 Å) to 1.940 Å (100 GPa, $Cmcm$, Cl---Cl 2.752 Å), respectively, while the interplane stacking Cl---Cl separation decreases from 4.024 Å to 2.514 Å. This effect leads to a band gap closure upon compression (Fig. 3d).

Our variable-composition search showed that above 160 GPa and below 251 GPa, the 1:1 stoichiometry is unstable and HCl will decompose into a mixture of $H_2Cl$ and $H_4Cl_7$. This founding is contradict to others research[10]. The tetragonal $P4/nmm$-HCl possesses a layered mackinawite-type structure, and becomes stable at pressures above 251 GPa. In this structure, adjacent HCl-based zigzag chains get close to form 2D-layers (Fig. 3c). The shortest interlayer Cl-Cl distance is 2.199 Å. In each layer, HCl zigzag chains run along the a- and b-axes. Here, H atoms have a tetrahedral coordination (Cl-H-Cl angle of 126.7°), while Cl atoms have a four-fold umbrella coordination, enabling the interlayer space to interact with lone electron pairs. The elongated H-Cl bond length reflects the increase in the number of H-Cl bonds under compression - four in $P4/mmm$ (1.574 Å at 300 GPa), two in symmetric $Cmcm$ HCl (1.442 at 100 GPa), one in molecular $Cmc2_1$ (1.313 at 0 GPa) - reflective of electron delocalisation in the covalent 2D layers. The $P4/nmm$ structure is metallic (Fig. 3d). Previous reported results proposed the following transition sequence

$Cmc2_1$ ->(35 GPa) $Cmcm$ ->(108 GPa) $P$-$1$ from *ab initio* evolutionary search[11]. This triclinic $P$-$1$ structure is always higher in energy than our predicted thermodynamically stable $Cmcm$ (39-160 GPa) and $P4/nmm$-HCl (251-500 GPa) structures.

*3.2. $H_2Cl$: HCl zigzag chains + $H_2$ units*

The hydrogen-rich compound $H_2Cl$ crystallizes in the monoclinic structure $C2/c$ (Fig. 5a). This structure is formed by planar symmetric zigzag HCl chains and $H_2$ dimers, with usual structural parameters along the L-shaped chains: H-Cl = 1.450 Å, H-Cl-H = 98.0º, and Cl-H-Cl = 179.4º at 100 GPa. $H_2$ dimers form planar zigzag chains with alternating short intramolecular (0.733 Å) and long intermolecular (1.563 Å, at 100 GPa) distances, as expected from band theory (Peierls distortion). The H-H-H angle is 142.1º. This encapsulated $(H_2)_n$ chain is insulating and presents a nice 'textbook' hydrogen chain model in an experimentally-synthesizable material. Pairs of stacked HCl chains are observed in the *ac* planes with an intra-pair separation of 0.901 Å and the shortest Cl--Cl separation of 2.496 Å compared to the van der Waals Cl-Cl separation of 3.50 Å at 0 GPa. Thus, some attractive interactions are in work in these paired chains, signature of a chain-pairing effect. The calculated DOS of $C2/c$ $H_2Cl$ (Fig. 5c) shows that this phase is an insulator in its pressure range of stability, i.e. 44 to 341 GPa.

At 341 GPa, the low-pressure $C2/c$ structure transforms to the rhombohedral $R$-$3m$ structure (Fig. 5b), which has a cubic close packing of Cl atoms in which all octahedral voids are filled by $H_2$ groups. At 450 GPa, the shortest H-Cl distance is 1.495 Å and the shortest intra- and interplanar Cl-Cl distances are 2.206 Å and 2.538 Å, respectively. The H-H distance (1.057 Å) is much longer than that of the $H_2$ molecule (0.74 Å), though still bonding, which indicates weakening due to depletion of the σ-bonding orbital that donates part of its electron density to the Cl framework. The band structure of the $R$-$3m$ phase is shown in Supplementary Fig. S2, and features 'flat band steep band' characteristics[31], which may suggest superconductivity. The electron-phonon coupling (EPC) calculation shows that the EPC constant λ is 0.66 at 450 GPa with an $\omega_{log}$ of 1368 K and $T_c$ = 43.9~44.8 K estimated using μ* = 0.1~0.15 in the Allen-Dynes modified McMillan equation[32].

*3.3. $H_3Cl$*

The first phase of the hydrogen-rich compound $H_3Cl$ phase crystallizes in the monoclinic $C2/c$ structure (Fig. 6a), which is stable in the pressure range 19-59 GPa. In this structure, the HCl units are packed in symmetric zigzag chains (H-Cl=1.519 Å, H-Cl-H=101.1º, at 25 GPa), between which non-bonded $H_2$ molecules are located - formally two $(H_2Cl_2)_n$ polymers and four $H_2$ units per unit cell. The presence of $H_2$ reduces the pressure of required to symmetrise the Cl-H-Cl chain in $Cmc2_1$-HCl. The intramolecular H-H distance is 0.741 Å (0.74 Å in gas phase $H_2$) and the intermolecular $H_2$-$H_2$ distance is quite long (1.781 Å). This interaction is clearly non-bonding with only van der Waals interactions. The $C2/c$ structure can be viewed as a host-guest inclusion compound: stacked 1D planar inorganic polymeric chains ($2H_2Cl_2$ per unit cell) with four encapsulated $H_2$ units in the van der Waals space. The enthalpy of formation from HCl and $H_2$ is calculated as -0.041 eV/atom at 50 GPa. An interesting phenomenon should be highlighted. In $C2/c$ $H_2$-HCl, the symmetrised ($H_2Cl_2$) chain is encountered at about 19 GPa, whereas hydrogen bond symmetrisation was observed in stoichiometric 1:1 HCl at a much higher pressure of ~39 GPa. Therefore, $H_2$ dimers act as either adding extra mechanical pressure or a chemical doping in the symmetrisation process, lowering the driving force of the phase transition due to the Peierls instability.

At 59 GPa, the monoclinic $C2/c$ phase transforms into an orthorhombic structure with the space group $P2_12_12_1$ (Fig. 6b) and then decomposes into $H_2Cl$ and $H_5Cl$ at 128 GPa. Free rotation is allowed around the bonding H-Cl chain, thus a non-planar form may exist. In the $P2_12_12_1$ phase (Z = 3), slightly helical zigzag HCl-chains run along the *b*-axis. The dihedral angle H-Cl–Cl-H is 14.1°. The H-Cl bond length is 1.442-1.444 Å and the H-Cl-H bond angle is 97.1° at 100 GPa. There are four $H_2$ dimers per unit cell located in the cavities between two slightly helical chains (H-H = 0.732Å at 100 GPa). The closest intermolecular H-H distance is 1.424 Å at 100 GPa. $P2_12_12_1$-$H_3Cl$ can be thought of as made of 1D helical HCl chains with delocalized covalent bonding and $H_2$ molecules. Its enthalpy of formation from HCl and $H_2$ is calculated as -0.07 eV/atom at 100 GPa.

Above 320 GPa, a $P2/m$ structure is stable (Fig. 6c). One can observe graphene-like arrays of alternant Cl and H atoms, some parallel and some perpendicular to the *ac* plane. The graphene-like nets are composed by the staggered packing of symmetric HCl zigzag chains. The H-Cl distance is 1.547 Å in the zigzag chains, while the H-Cl length when connected via two zigzag chains is 1.473 Å. The interspacing between two graphene-like nets distances is 1.839 Å at 400 GPa. The shortest H-Cl distance is 1.438 Å at 400 GPa, typical of the values for bridging $H_2$ molecules to graphene-like nets. $H_2$ units are observed in channels formed by graphene-like nets. The H-H distances are 0.779 Å arranged alternately along the *c*-axis. The band structure calculations reveal that the $C2/c$ and $P2_12_12_1$ phases of $H_3Cl$ are indirect-gap insulators with DFT band gap of 4.68 eV and 4.22 eV, respectively, while the $P2/m$ phase is metallic (See Fig. 6).

### 3.4. $H_5Cl$: the existence of $H_3^+$ units

At 106 GPa, the hydrogen-rich compound $H_5Cl$ becomes stable. Its lowest-pressure structure *Pc* (Fig. 7a) contains planar chains composed of triangular $H_3$ units connected to two chlorine atoms (H-H = 0.864-0.913 Å, H-Cl = 1.605-1.680 Å at 150 GPa). The structure also contains isolated $H_2$ dimers (H-H = 0.748 Å), $H_3$ groups and Cl atoms. Formally, the $H_3$ unit may be viewed as the well-known $H_3^+$ cation interacting with one $Cl^-$ anion.

The $H_3^+$ cation deviates slightly from equilateral triangle, the shape experimentally and theoretically observed in gas phase $H_3^+$, the smallest and simplest aromatic molecule. The H-H bond lengths are elongated (H-H = 0.864-0.913 Å) by 19% relative to the calculated distance in encapsulated $H_2$ dimers in the *Pc* phase (0.748 Å) or the experimental free gas phase $H_2$ distance (0.74 Å). Such elongation is expected for an electron-deficient system (Fig. 7c). The molecular orbital diagram for symmetric ($D_{3h}$) $H_3^+$ is reproduced in Fig. 7d. The stability of $H_3^+$ is due to occupation of the a'$_1$ bonding MO and the large HOMO a'$_1$-LUMO e' gap. The formal H-H bond order is 1/3 per H-H link (1 in $H_2$), explaining the longer hydrogen-hydrogen distances for $H_3^+$ ($H_3^+$ may be described by the well-known three-center two-electron bonding scheme). $H_3^+$ is stabilised in this solid-state compound by electrostatic interactions with the anionic chloride network.

$H_2$ can act as an electron donor in combination with a proton to yield $H_3^+$. By analogy, $H_2$ may also combine with $H_3^+$, leading to the higher homolog $H_5^+$. The existence of $H_3^+$ and $H_5^+$ have been firmly established by mass spectrometry and quantum-mechanical calculations[33-36]. The $H_5^+$ ion appears in the high-pressure phase of $H_5Cl$ with the space group $P2_1/c$. The H-H distances reveal the existence of the planar $H_5^+$ unit: all H-H distances are in the range of 0.753-1.085 Å at 350 GPa. In the psudo-equilateral $H_3$ unit, the H-H bond lengths are 0.856 Å to 0.880 Å at 350 GPa. Two H-H bonds are significantly longer (1.085 Å) than the H-H bonds reported in $H_3$ units, indicating weaker bonds. The shortest distance is 0.753 Å, similar to the length of the H-H covalent bond in free $H_2$. Therefore, this planar $H_5$ unit may be viewed as $H_3^+$ interacting with a slightly activated $H_2$ dimer. A five-center four-electron bonding scheme (5c-4e) is assigned: such a bonding mode has previously been observed in protonated hydrogen-based clusters[37], the global-minimum structure of which is, however, non-planar. The calculated DOS of $H_5Cl$ (Fig. S2) shows that this compound is an insulator upto 479 GPa. Obviously, H-rich system is not easily metallised compare to Cl-rich system.

*3.5. $H_4Cl_7$: chlorine-based Kagomé layers intercalated with zigzag HCl chains*

Besides hydrogen-rich H-Cl compounds, we also uncovered a chlorine-rich compounds $H_4Cl_7$, which has never been reported before. $C2/m$-$H_4Cl_7$ was found to be stable in the pressure range 90-278 GPa, and it transforms to another monoclinic structure with the space group $C2$ at 278 GPa (Fig. 2). In the low-pressure $C2/m$ phase, three different networks can be found. The first is two $(HCl)_2$ zigzag chains intercalated between planar Kagomé covalent layers ($4H_2Cl_2 + 2Cl_3 = H_8Cl_{14}$). The second is a 1D zigzag chain, which has the topology encountered in the $Cmcm$-HCl phase with two-coordinate bent chlorine atoms and two-coordinate linear hydrogen atoms (H-Cl = 1.46-1.47 Å; H-Cl-H = 99.3°-100.1°, Cl-H-Cl = 174.1°, 100 GPa). The atomic arrangement can be rationalised by the VSEPR model (Fig. 8a). The final network is Cl-Cl zigzag chains running along the *b*-axis with a bond length of 2.563 Å at 100 GPa.

The planar Kagomé $(Cl_3)_2$ covalent layers possess four-coordinate rectangular planar chlorine centres with Cl-Cl bond distances of 2.236-2.273 Å and bond angles of 60.5° and 119.5° at 100 GPa. Seven electrons may be formally assigned to each chlorine centra, thus Cl is the $AX_4E_{3/2}$ type, a rectangular planar $AX_4$ unit (one electron less than that of $XeF_4$)[3]. The zigzag chains are oriented in such a way that the chlorine atoms are located in the six- and three-membered ring-based channels made by the stacked Kagomé layers (Fig. 8a).

At 278 GPa, the $C2/m$ phase slightly distorts to the lower-symmetry $C2$ phase with more complex substructures. Between these flat Kagomé layers, there are two zigzag $(HCl)_2$ chains with two types of Cl atoms: one is coordinated to two bridging H atoms (H-Cl = 1.367-1.387 Å), the other is four-coordinated to two bridging H atoms (H-Cl = 1.375-1.379 Å) and two Cl atoms through short and long bonds (Cl-Cl = 2.051-2.223 Å). Even if this phase displays delocalized bonding, see the Cl--Cl long distance of 2.22 Å, signature of a deficient covalent bond, their local structures are rationalised using VSEPR rules (Fig. 9c) from single Lewis resonant structures. Formally, the two-coordinated Cl atoms are $AX_2E_2$ with a bent geometry (H-Cl-H = 99.2°). Four-coordinated Cl atoms are $AX_4E_1$ with a local butterfly-like structure (H-Cl-H = 95.2°, Cl-Cl-Cl = 173.1°). Moreover, the two-coordinate H ion is $AX_2E_0$ (linear). Finally, one can see linear chains of Cl atoms with alternating inter-chain distances (2.051 Å and 2.223 Å, at 300 GPa) running through the centre of $Cl_6$ hexagons. The $C2$ structure is stable up to 445 GPa; at higher pressures it decomposes into HCl and Cl. Figures 8b and 9b shows the electronic density of states for $C2/m$-$H_4Cl_7$ and $C2$-$H_4Cl_7$, respectively. We find that both $C2/m$ and $C2$ phases are metallic contributed by the Cl-*p* states.

**4. Conclusions**

In summary, we have extensively explored the stable compounds and their structures in the H-Cl system up to 500 GPa. Besides well-known HCl, we also discovered several H-rich compounds, $H_2Cl$, $H_3Cl$ and $H_5Cl$, and a Cl-rich compound $H_4Cl_7$. At ambient pressure and low temperature, $H_2$, $Cl_2$ and HCl molecular crystals are formed by weak intermolecular van der Waals interactions. HCl discontinuously exits below 160 GPa ($Cmcm$) and above 251 GPa ($P4/nmm$), while it dissociates into

$H_2Cl$ (*C2/c*) and $H_4Cl_7$ (*C2/m*) in between.

Under pressure, HCl reacts with $H_2$ to yield stable $H_3Cl$ stoichiometry with $H_3^+$ and $Cl^-$ ions. Further reaction with $H_2$ occurs under pressure, producing $H_5Cl$, where $H_2$ combines with $H_3^+$ and the higher homolog $H_5^+$ emerges in $H_5Cl$. Chlorine-based Kagomé layers (indices 3636) of Cl atoms intercalated by zigzag HCl chains are found in the chlorine-rich hydrides $H_4Cl_7$. These findings of unexpected complexity in the simple H-Cl system will be helpful in understanding the effect of pressure on chemical bonding.

**Acknowledgements**

We thank the National Natural Science Foundation of China (Grants No.51372203 and No.51332004), the Foreign Talents Introduction and Academic Exchange Program (Grant No.B08040), DARPA (Grants No.W31P4Q1310005 and No.W31P4Q1210008), Poitiers University, the CNRS, and the Government of the Russian Federation (Grant No.14.A12.31.0003) for financial support. The authors also acknowledge the High Performance Computing Center of NWPU for allocation of computing time on their machines.

**Author Contributions**

A.R.O. initiated the project. Q.-F.Z., S.-Y.Y. and D.L. performed the structure prediction and electronic structure calculations. D.L. and S.-Y.Y. performed superconductivity calculation. G. F. S.-Y.Y. and A.R.O performed the bonding analysis. S.-Y.Y., D.L. and G. F. prepared all the figures and tables. Q.-F.Z., S.-Y.Y., D.L. and G.F. wrote the manuscript text. All authors discussed the results and made comments to the manuscript.

**Additional information**

Competing financial interests: The authors declare no competing financial interests.

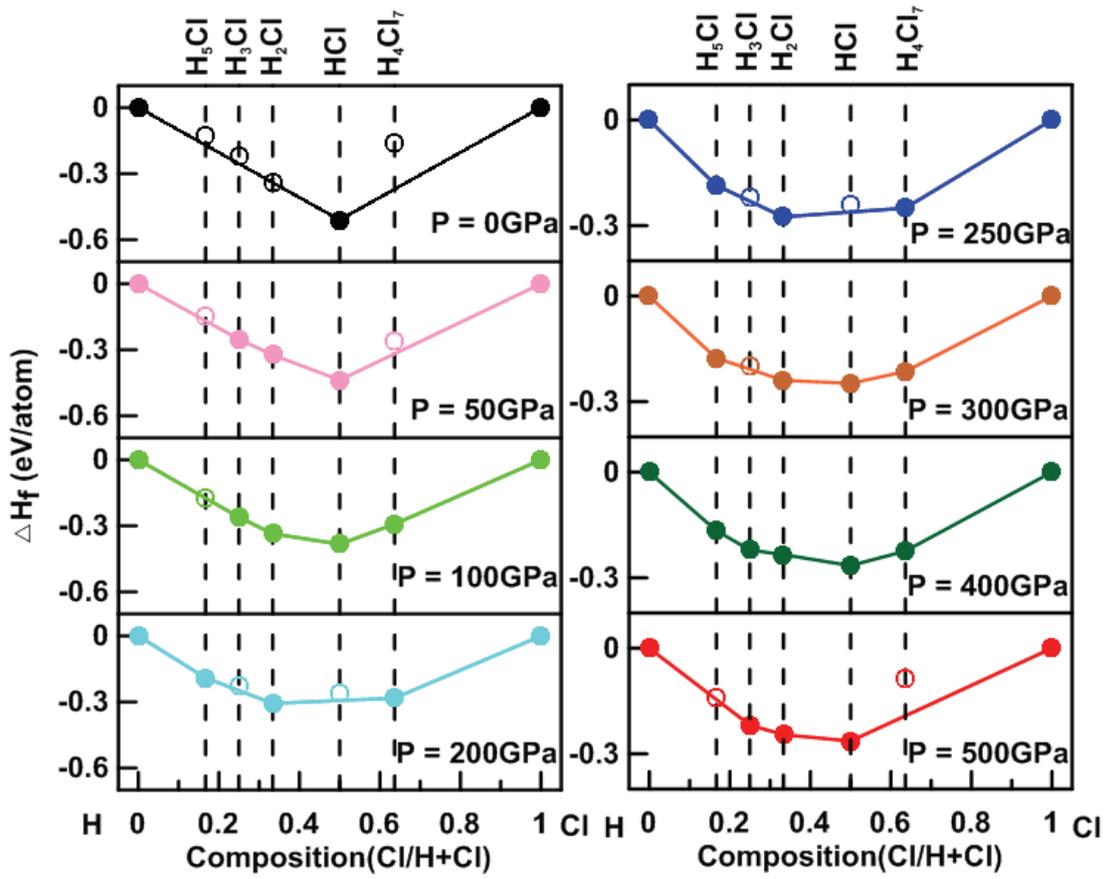

Fig. 1 (Color online) Convex hull diagrams for the H-Cl system over the pressure range 0-500 GPa. Enthalpies of formation (per atom) of the $H_xCl_y$ phases are calculated with respect to elemental chlorine and hydrogen in their corresponding most stable phases. $H_xCl_y$ structures that are stable with respect to disproportionation are those that form the convex hull (solid circles: stable; open circles: metastables).

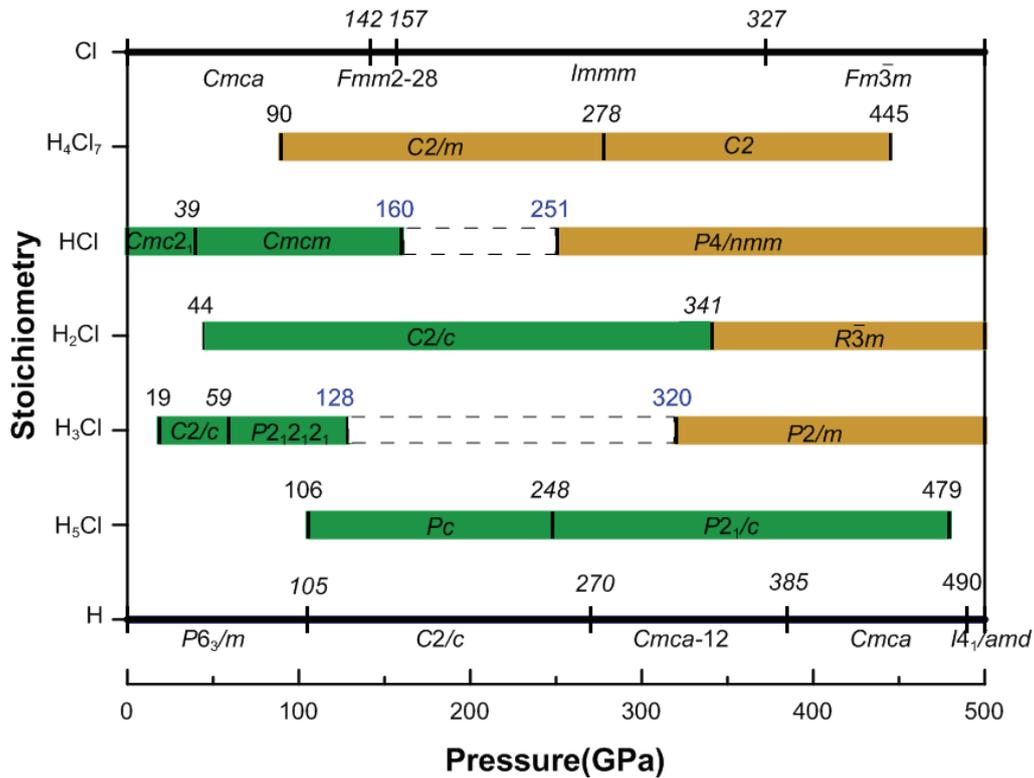

Fig. 2 (Color online) Pressure-composition phase diagram of the H-Cl system. The blue pressure values indicate the initial pressure at which the compound becomes unstable with respect to disproportionation into other hydrogen chlorides or chlorine and hydrogen (dashed line blocks). The phase transition pressure is indicated in italics. Pressures are given in GPa. (The green blocks indicate the insulators, and the brown blocks indicate metallics)

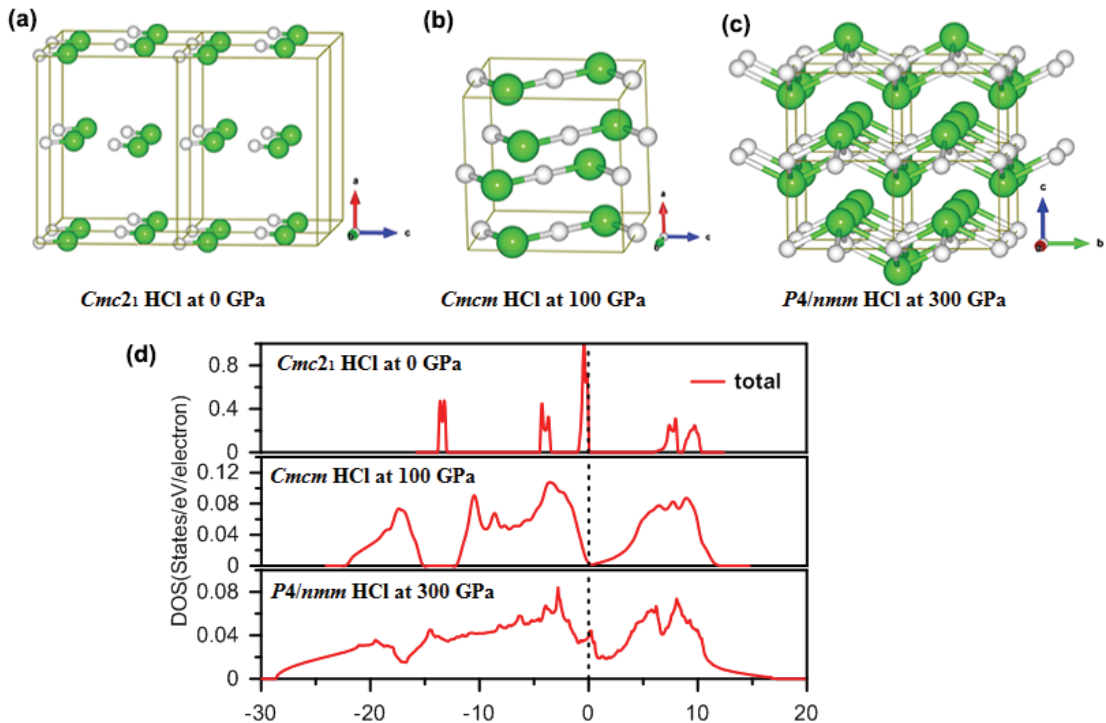

Fig. 3 (Color online) Crystal structures of HCl. (a) $Cmc2_1$ phase at 0 GPa; (b) $Cmcm$ phase at 100 GPa; (c) $P4/nmm$ phase at 300 GPa; (Cl = green large sphere, H = white small sphere) (d) The calculated

total electronic density of states (DOS) for *Cmc2$_1$*, *Cmcm* and *P4/nmm* phases, respectively. The energy of the Fermi level has been set to zero. For partial DOS calculations, see the Fig. S2.

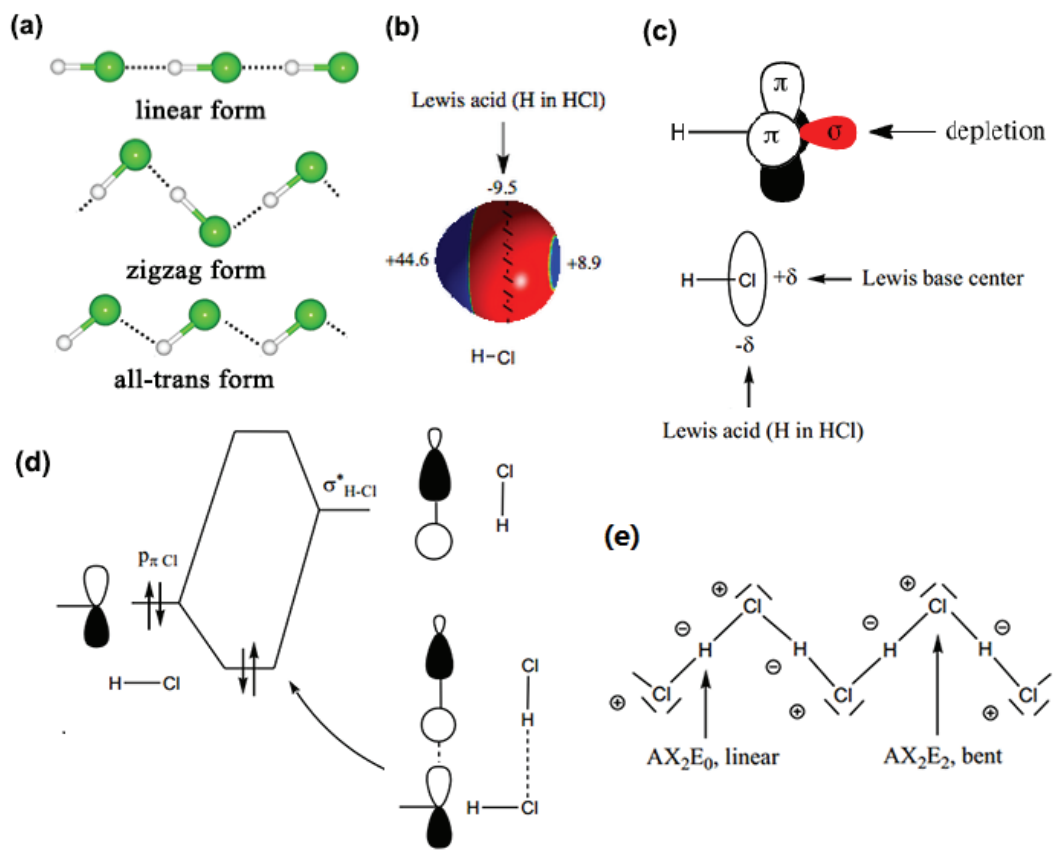

Fig. 4 (Color online) (a) 1D planar inorganic HCl chain in linear, zigzag (L-shaped) and all-trans forms; (b) Computed electrostatic potential $V_S(r)$, B3PW91/aug-cc-pVTZ, on the 0.001 au molecular surface of HCl (positive in blue, negative in red). For HCl, the locations and of the most positive ($V_{S, max}$) and negative ($V_{S, min}$) potentials are indicated by black hemispheres and black crosses, respectively; their values, in kcal/mol, are indicated. (c) schematic σ-hole concept for halogen and hydrogen bonds; (d) Orbital interaction diagram of HCl---HCl dimer in a L-shaped coordination; (e) A Lewis structure for 1D zigzag HCl-chain with assigned formal charges and VSEPR AX$_n$E$_m$ notation.

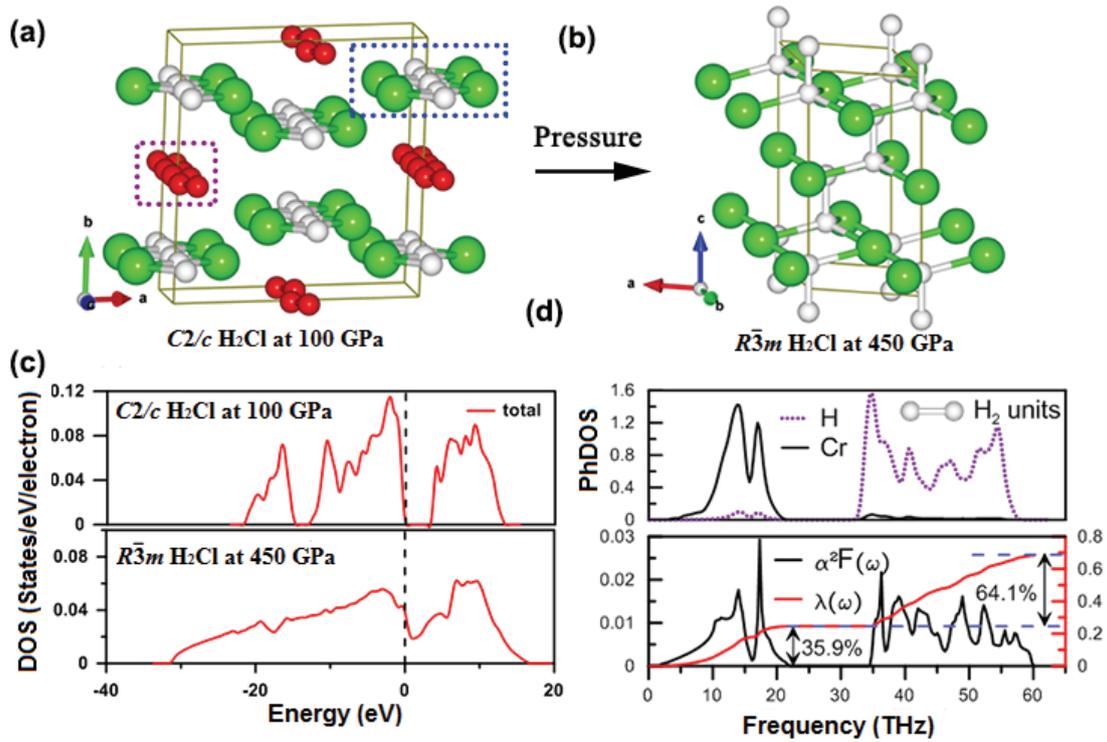

Fig. 5 (Color online) Crystal structures of $H_2Cl$ with the space group of $C2/c$ (a) and $R\bar{3}m$ (b). Green spheres represent chlorine atoms while red and white spheres represent hydrogen atoms. (c) Calculated total electronic density of states for $H_2Cl$. (d) Partial phonon density of states (PhDOS) for the $R\bar{3}m$ phase and Eliashberg phonon spectral function $\alpha^2F(\omega)$ and electron-phonon integral $\lambda(\omega)$ as a function of frequency at 450 GPa.

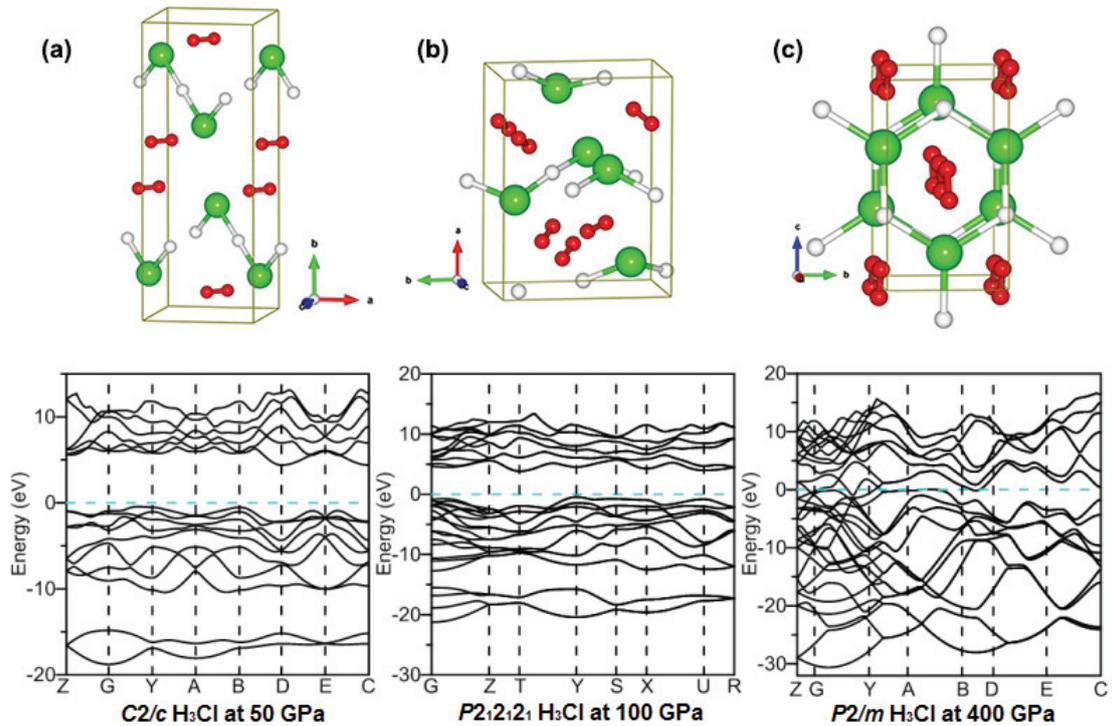

Fig. 6 (Color online) Crystal structures and corresponding band structures of the $H_3Cl$ phases. (a) $C2/c$, (b) $P2_12_12_1$, and (c) $P2/m$. The large spheres represent chlorine atoms and small spheres represent

hydrogen atoms. H$_2$ units are depicted in red.

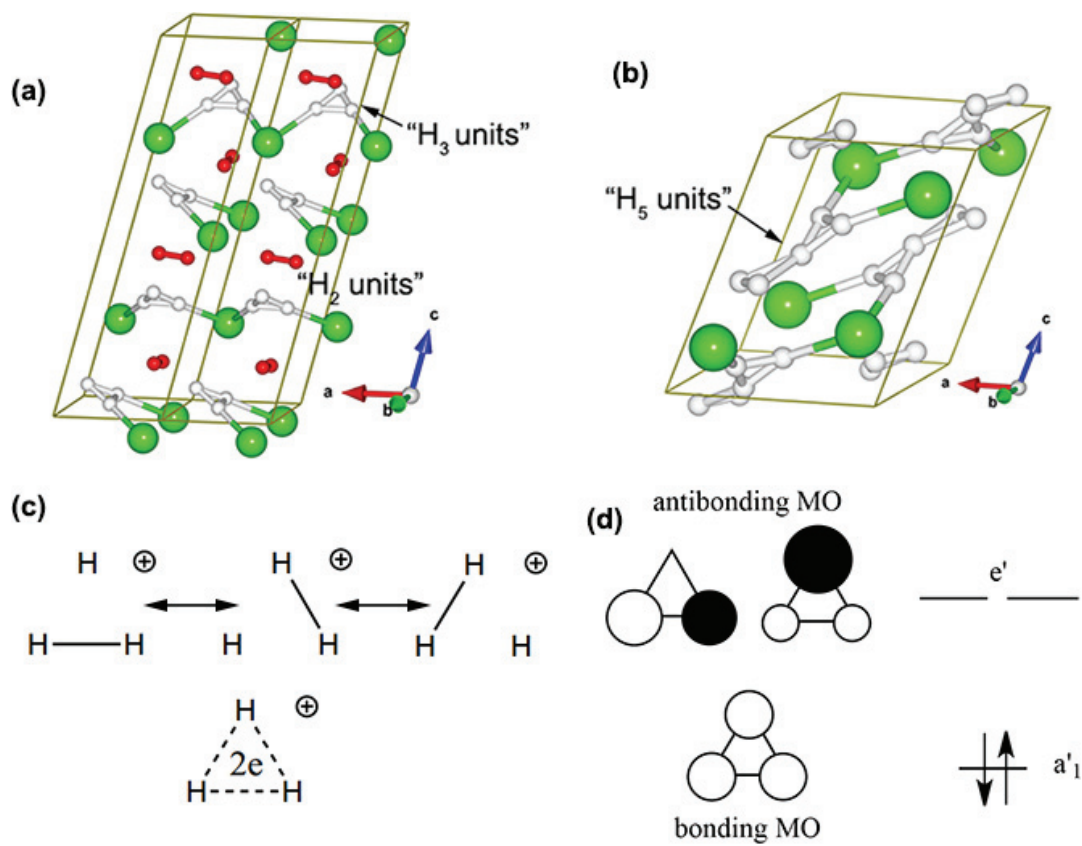

Fig. 7 (Color online) (a, b) Crystal structures of H$_5$Cl; H$_2$ units are depicted in red while H$_3$ and H$_5$ units are shown in white; (c) Lewis structures for the H$_3^+$ cation (3c-2e bonding); (d) molecular orbital (MO) diagram for the equilateral H$_3^+$ cation ($D_{3h}$ point group).

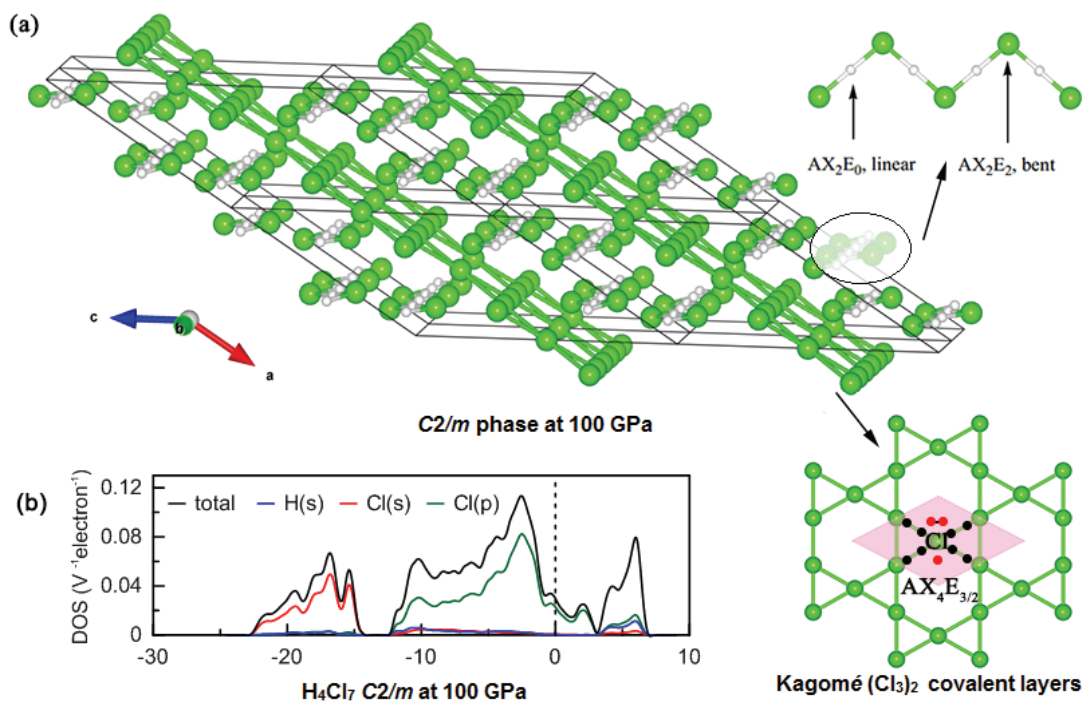

Fig. 8 (Color online) (a) Crystal structures of $C2/m$ H$_4$Cl$_7$ with VSEPR analysis of the H-Cl zigzag

chains and the planar Kagomé $(Cl_3)_2$ covalent layers;(b) the calculated total and partial density of states for $C2/m$ $H_4Cl_7$.

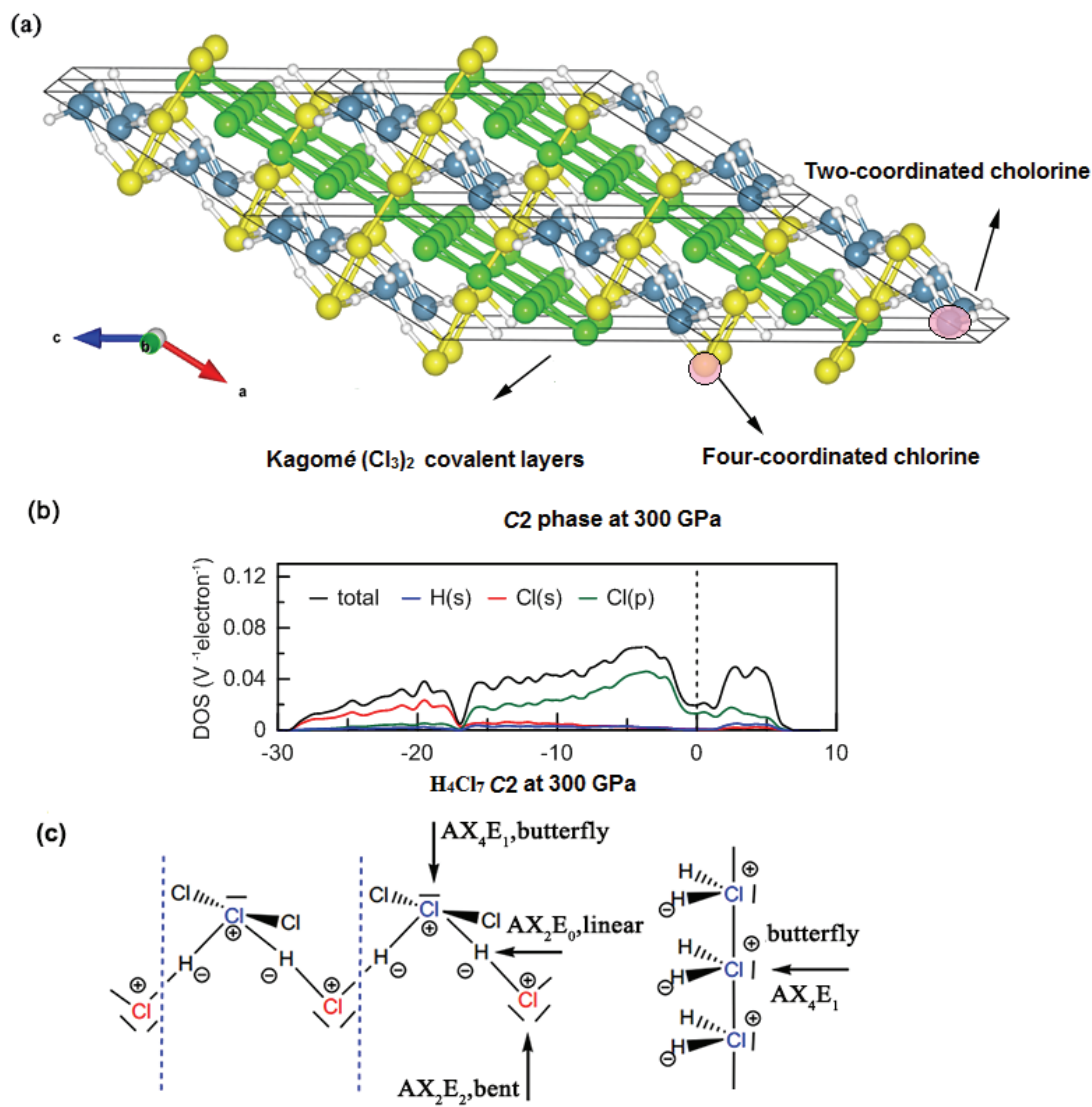

Fig. 9 (Color online) Crystal structures of $C2$ $H_4Cl_7$ with VSEPR analysis of the planar Kagomé $(Cl_3)_2$ covalent layers; (b) Calculated total and partial density of states for $C2$ $H_4Cl_7$.